\documentclass[9pt,twocolumn,twoside]{revtex4-1}
\usepackage{bm}
\usepackage{csquotes}
\usepackage{mathtools}
\usepackage{textcomp}
\usepackage{amsmath}
\usepackage{amssymb}
\usepackage[utf8]{inputenc}
\usepackage{xcolor}
\usepackage[normalem]{ulem}
\usepackage{amsthm}
\usepackage{enumitem}
\theoremstyle{remark}
\DeclareFontFamily{U}{wncy}{}
\DeclareFontShape{U}{wncy}{m}{n}{<->wncyr10}{}
\DeclareSymbolFont{mcy}{U}{wncy}{m}{n}
\DeclareMathSymbol{\Sha}{\mathord}{mcy}{"58}
\usepackage[pretty]{revquantum}
\begin{document}
	\title{Self-sustained Molecular Rectification without External Driving or Information}
	
	\author{Jiantang Jiang}\thanks{Corresponding author: Jiantang Jiang (email: tumport@126.com)} 
	\affiliation{Zhongshan Shiruan Software Technology Co., Ltd., Zhongshan City, Guangdong Province 528429, China
	}
	\begin{abstract}
		Rectifying thermal white noise into directed motion is generally believed to require the consumption of energy or information, as exemplified by Maxwell’s demon–type feedback controllers. Here we demonstrate a molecular rectification mechanism that operates without any external energy or information flow. An ion-induced asymmetry between two liquid–vapor interfaces creates unequal surface barriers, enabling the harvesting and redistribution of surface energy released during condensation. Molecular dynamics simulations show that this intrinsic kinetic asymmetry sustains a persistent net water flux. Our results suggest that asymmetric potential energy landscape alone can rectify thermal fluctuations, revising the conventional understanding of noise-driven transport. 
	\end{abstract}
	
	\maketitle
	
	\section{Introduction}
	\label{sec:1}
	Since Maxwell’s seminal thought experiment\cite{01}, the rectification of thermal fluctuations has been intimately connected to information processing and energy consumption. Experimental realizations of Maxwell's demon–type feedback controllers have been demonstrated across a wide range of platforms\cite{02,03,04,05,06,07,08,09}. Extensive studies of biological\cite{10,11,12,13,14} and synthetic\cite{15,16,17,18,19,20,21,22} ratchets have reinforced this paradigm: asymmetric potentials or geometric constraints, in the absence of energy or information input, are insufficient to induce a nonzero probability current. Directed transport is therefore generally attributed to time-dependent driving, feedback control, or nonequilibrium chemical reactions. Whether purely static asymmetries can induce directed transport without invoking an external resource remains an open and fundamental question.
	
	In this Letter, we demonstrate a molecular rectification mechanism driven solely by thermal white noise. An ion concentration difference between two liquid–vapor interfaces generates unequal surface barriers, leading to an intrinsic kinetic asymmetry. Energy released during condensation at one interface is harvested via surface contraction and transferred to facilitate barrier crossing at the opposite interface, sustaining a net flux without external energy or information consumption. Our results identify an interfacial energy–harvesting mechanism that enables autonomous rectification and provides a new perspective on noise-driven transport.
	
	\section{Model and methods}
	\label{sec:2}
\begin{figure}[t]
    \centering
    \includegraphics[width=0.7\columnwidth]{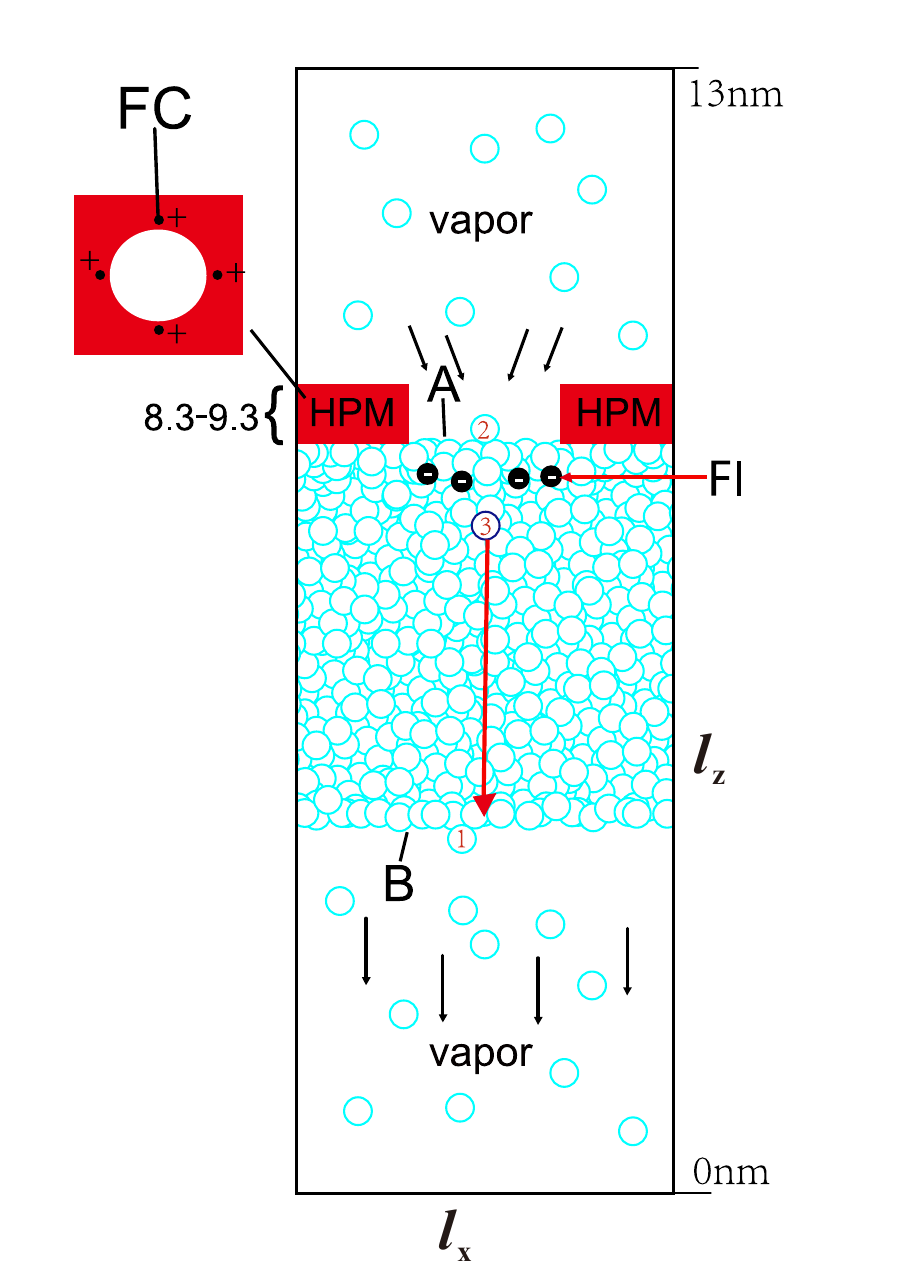}  
    \caption{Simulation setup. HPM, hydrophobic porous membrane; FC, fixed charges on the pore sidewall; FI, free ions in water. The upper-left inset shows the top view of the HPM. The figure is not drawn to scale.}
	\label{fig:1}
\end{figure}

	The simulation setup is shown schematically in Fig.~\ref{fig:1}. A simulation box (dimensions: $l_x = l_y = $ 2.6 nm and $l_z = $ 13 nm contains a 4-nm-thick water slab (950 molecules) centered along the z direction. A 1-nm-thick hydrophobic porous membrane (HPM) with a central pore of diameter 1.6 nm is placed on top of the slab and held fixed. Four fixed charges, all lying in the same xy plane, are attached to the pore sidewall. Four free ions of opposite charge disperse in the water, ensuring overall charge neutrality. Two liquid–vapor interfaces are formed: interface A at the pore opening and interface B at the free lower surface.

    Periodic boundary conditions are applied in all directions, and the system is coupled to a thermal bath (see Secs. I and II of Supplemental Material (SM)\cite{23} for details). Directed transport is quantified by the net number of water molecules (NNWM) crossing the plane $z = 0$. A monotonic temporal increase of NNWM indicates a sustained vapor flux from surface B to surface A.

\begin{figure}[t]
    \centering
    \includegraphics[width=1.0\columnwidth]{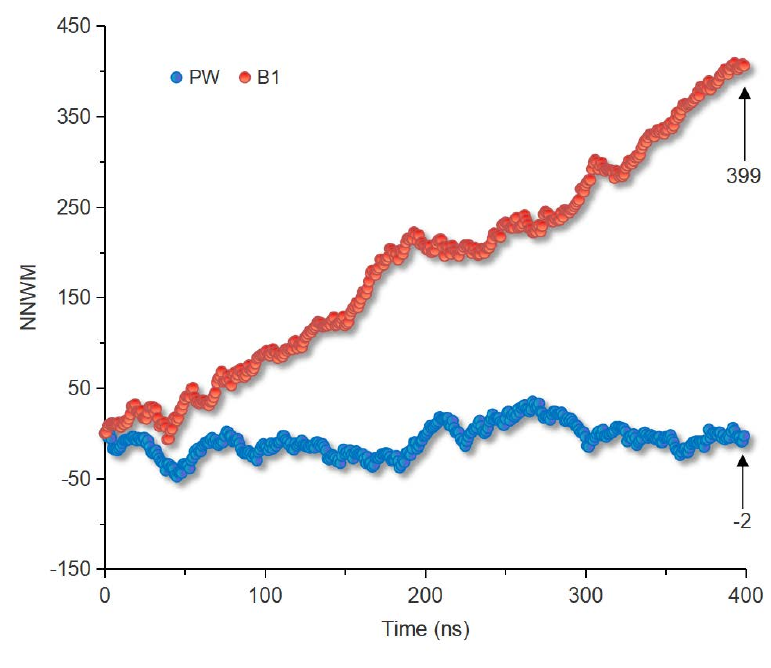}  
    \caption{Variation of NNWM values as a function of time in simulations of PW and B1. }
	\label{fig:2}
    \end{figure}

	In a representative simulation (B1), fixed charges are $Cl^-$ ions and free ions are $Na^+$. The system is simulated for 400 ns at 370 K. As shown in Fig.~\ref{fig:2}, NNWM exhibits a clear directional trend, demonstrating persistent transport. The transport cycle consists of evaporation at surface B, vapor migration through periodic boundaries, condensation at surface A, and liquid return flow from A to B. The control simulation composed of only the pure water slab and the HPM, obtained by removing all the fixed charges and free ions, shows no directed flux (PW in Fig.~\ref{fig:2}), confirming the essential role of ion-induced asymmetry. Sec. III in SM\cite{23} provides more simulations and details.

	\section{Intrinsic kinetic asymmetry}
	\label{sec:3}
	The ion concentration difference between the surfaces $A$ and  $B(X_{ion,A}>X_{ion,B})$ renders the liquid slab inhomogeneous. For an ideal solution, the chemical potential $\mu$ of a component is related to its mole fraction $X$ by
	\begin{equation}\label{eq:1}
		\mu = \mu ^\circ + R T \ln X 
	\end{equation}
	where $\mu ^\circ$ is the chemical potential of the component at standard state, $R$ is the gas constant, $T$ is the temperature. Since the liquid slab consists only of water and free ions, the mole fractions of ions and water molecules at the two liquid surfaces satisfy
	\begin{equation}\label{eq:2}
		1 = X_{ion,A} + X_{water,A} = X_{ion,B} + X_{water,B}
	\end{equation}

	Because $X_{ion,A} > X_{ion,B}$, it follows that $X_{water,B} > X_{water,A}$. Substituting into Eq.~\eqref{eq:1}, the chemical potentials of water molecules of the two liquid surfaces and the difference between them can be obtained as 
	\begin{equation}\label{eq:3}
		\mu_{diff} = \mu_{water,B} - \mu_{water,A} > 0
	\end{equation}

	The chemical potential gradient $\mu_{water,B} > \mu_{water,A}$ implies a difference in the saturated vapor pressures of the two liquid surfaces, namely $P_b > P_a$, consistent with Raoult’s law and confirmed by the virtual-wall model (Sec. IV of SM\cite{23}). From Eqs.~\eqref{eq:1}–~\eqref{eq:3}, one further obtains
	\begin{align}\label{eq:4}
	\mu_{diff} &= R T \ln (X_{water,B}/X_{water,A})\nonumber\\
	&= R T \ln [(1-X_{ion,B})/(1-X_{ion,A})]
    \end{align}

\begin{figure}[t]
    \centering
    \includegraphics[width=1.0\columnwidth]{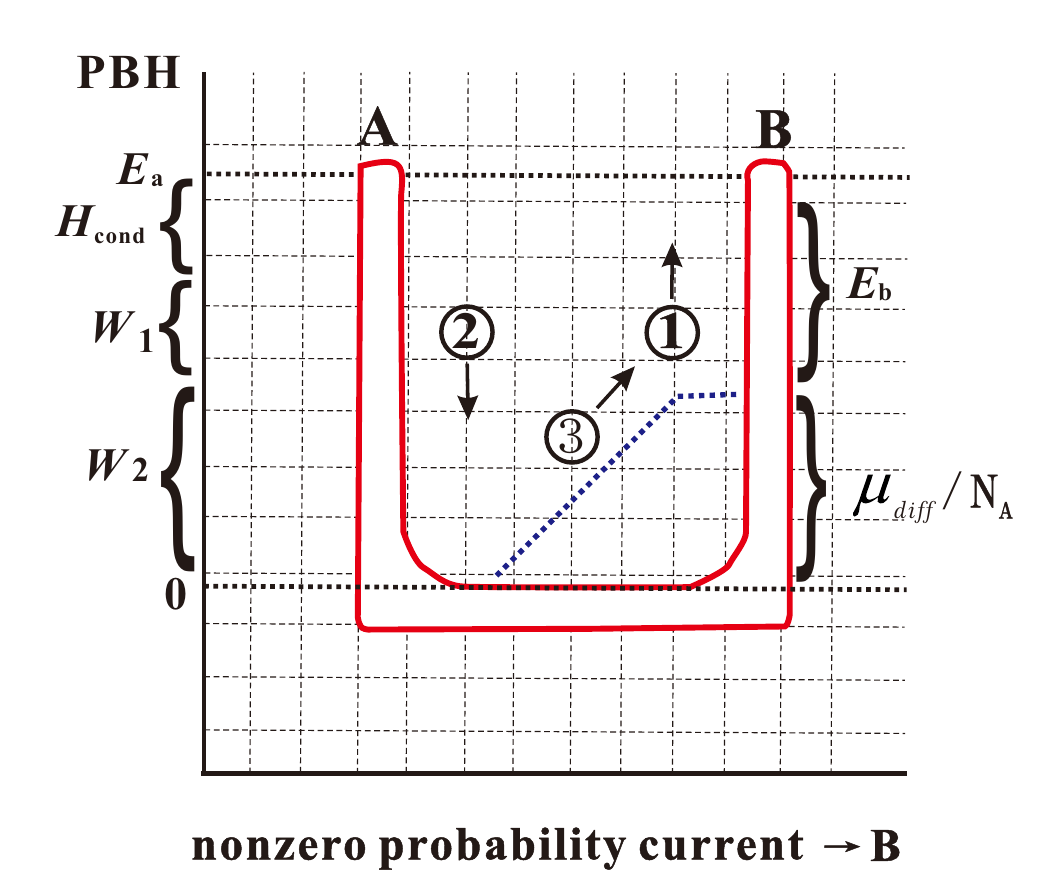}  
    \caption{Liquid molecules trapped in a potential well composed of surface barriers A and B. PBH denotes the potential barrier height. The zero of potential energy is defined at the well bottom.}
	\label{fig:3}
    \end{figure}

	At the microscopic level, incomplete screening of ionic electric fields at the liquid surface introduces an additional surface potential\cite{27}. Experimentally, inorganic ions have been shown to suppress water evaporation\cite{28}, as strong ion–water interactions hinder the librational motion of interfacial water molecules\cite{29}.Therefore, compared to liquid surface B, liquid surface A has a higher barrier $(E_a > E_b)$ and a lower evaporation rate. The barrier height difference equals the chemical potential difference of water molecules between the two liquid surfaces, such that
	\begin{equation}\label{eq:5}
		E_a = E_b + \mu_{diff} / N_A
	\end{equation}
	where $N_A$ is Avogadro’s number. From Eq.~\eqref{eq:4}, one further obtains
	\begin{equation}\label{eq:6}
		E_a - E_b = R T \ln [(1-X_{ion,B})/(1-X_{ion,A})] / N_A
	\end{equation}
	That establishes a direct connection between the barrier height difference and the ion concentrations of the two liquid surfaces. Water molecules in the slab can be viewed as confined particles in an asymmetric potential well formed by the two surface barriers (Fig.~\ref{fig:3}). In Figs.~\ref{fig:1} and ~\ref{fig:3}, the same numbers indicate the same molecules. When the system reaches a stationary state with a stable flux, evaporation from surface B requires energy $E_b$, equal to the latent heat per molecule; i.e.,
	\begin{equation}\label{eq:7}
		H_{vap} = E_b
	\end{equation}

	Molecule 1 is a water molecule evaporating from liquid surface B. During its escape, its surface area is maximized, gaining a surface energy increment equal to $E_b$. Molecule 2 is a vapor molecule falling onto liquid surface A. At the instant of contact, the area of liquid surface A increases and it subsequently contracts spontaneously to minimize its surface energy. The surface contraction induced by the condensation of each molecule displaces all liquid molecules slightly toward surface B, which is effectively equivalent to driving a single molecule (molecule 3) from surface A to surface B. This constitutes a process of surface-energy harvesting, and the total work performed consists of two parts. $W_1$ is dissipated by viscous friction. $W_2$ raises the chemical potential of molecule 3 by $\mu_{diff} / N_A$; i.e., $W_2 = \mu_diff/N_A$.With this mechanism, liquid surface A will not get diluted despite continuous condensation on it.

	$W_2$ indirectly drives the vapor since the vapor flux is driven by the chemical potential difference. Therefore, the entire transport cycle is driven solely by surface contraction. Although the system exchanges heat with the thermal bath, the net heat transfer averages to zero. Following energy conservation, from Eq. \eqref{eq:7} and Fig.~\ref{fig:3} one obtains
	\begin{equation}\label{eq:8}
		E_a = H_{vap} + \mu_{diff} / N_A = W_1 + W_2 + H_{cond}
	\end{equation}
	where $H_{cond}$ is the latent heat of condensation per molecule. Both $H_{cond}$ and $H_{vap}$ are taken to be positive as they characterize energy scales rather than energy flows. With this energy-harvesting and transfer mechanism, the molecules at liquid surface B are always at a higher potential-energy state and thus have a higher escape probability, manifesting as a persistently higher saturated vapor pressure at surface B, which sustains the directed flux. The energy harvested per molecule condensing on surface A is $W_1 + W_2$. Since $E_a > (W_1 + W_2)$, the surface contraction can proceed spontaneously. Increasing the ion concentration difference deepens the effective potential well and enhances rectification. Frictional and latent-heat dissipation render the process statistically irreversible.

	Unlike conventional ratchets—such as feedback-based\cite{04,05}, flashing \cite{30,31}, rocking\cite{32}, correlation-driven\cite{33}, or geometric ratchets\cite{22,34}—which require external energy or information to break kinetic symmetry\cite{35,36}, our present mechanism is autonomous. The "gambling demon"\cite{37}, proposed by Gonzalo et al., may occasionally acquire more free energy than the work it consumes; however, such performance lacks sustainability. A long simulation up to 4 $\mu s$ yields NNWM=4,352, confirming sustainability of directed transport of our model (Sec. V of SM\cite{23}).

	\section{Quantifying the Asymmetry via Relative Entropy}
	\label{sec:4}
	The separate equilibrium vapor phases A and B are at the same temperature but differ in number density (Sec. IV in SM\cite{23}). Treating them as ideal gases, the chemical potential difference is related to the relative entropy as\cite{38}
	\begin{equation}\label{eq:9}
		D_{KL(P\| Q)} \cdot  N_A = \mu_{diff} / (kT)
	\end{equation}
	where $D_{KL(P\| Q)}$ is the relative entropy between the single-particle phase-space distributions of the two vapor phases. Combining with Eq.\eqref{eq:4}, one obtains
	\begin{equation}\label{eq:10}
		D_{KL(P\| Q)} \cdot  N_A = R T \ln [(1-X_{ion,B})/(1-X_{ion,A})] / (kT)
	\end{equation}
	Simplifying the above expression yields the KL divergence per particle as
	\begin{equation}\label{eq:11}
		D_{KL(P\| Q)} = \ln [(1-X_{ion,B})/(1-X_{ion,A})] 
	\end{equation}

	Thus, the degree of kinetic asymmetry is directly linked to the ion concentrations of the two liquid surfaces and can also be quantified by relative entropy. Kurchan established the relation between entropy production and relative entropy\cite{39}, and R. Kawai et al. derived the relation between average dissipated work and relative entropy\cite{40}. Our work can be connected to current thermodynamic frameworks via relative entropy. Within theoretical frameworks such as the generalized Jarzynski equality\cite{41} and the Sagawa–Ueda inequality\cite{42}, information is recognized as a resource for extracting positive work, but the acquisition or erasure of information incurs a thermodynamic cost constrained by the Landauer principle\cite{43}. The key feature of our model is that kinetic asymmetry alone suffices to rectify stochastic motion, without consuming any external energy or information. Our model is feasible using existing nanofabrication techniques and is amenable to development as a controllable nanofluidic device (Sec. VI in SM\cite{23}).

	\section{Conclusion}
	\label{sec:5}
	Our results demonstrate that thermal fluctuations can be rectified without external driving or information processing, revising the prevailing view that asymmetric potentials alone are insufficient to generate directed motion. The most fundamental difference between our model and other ratchet systems is the presence of an intrinsic energy-harvesting and transfer mechanism, which will provide valuable insights for future studies.

\end{document}


\preprint{APS/123-QED}

\title{Supplemental Material for the manuscript "Self-sustained Molecular Rectification without External Driving or Information"}

\maketitle

\section{The real system}\label{supp_sec:1}
\begin{figure}[H]
    \centering
    \includegraphics[width = 0.5\linewidth]{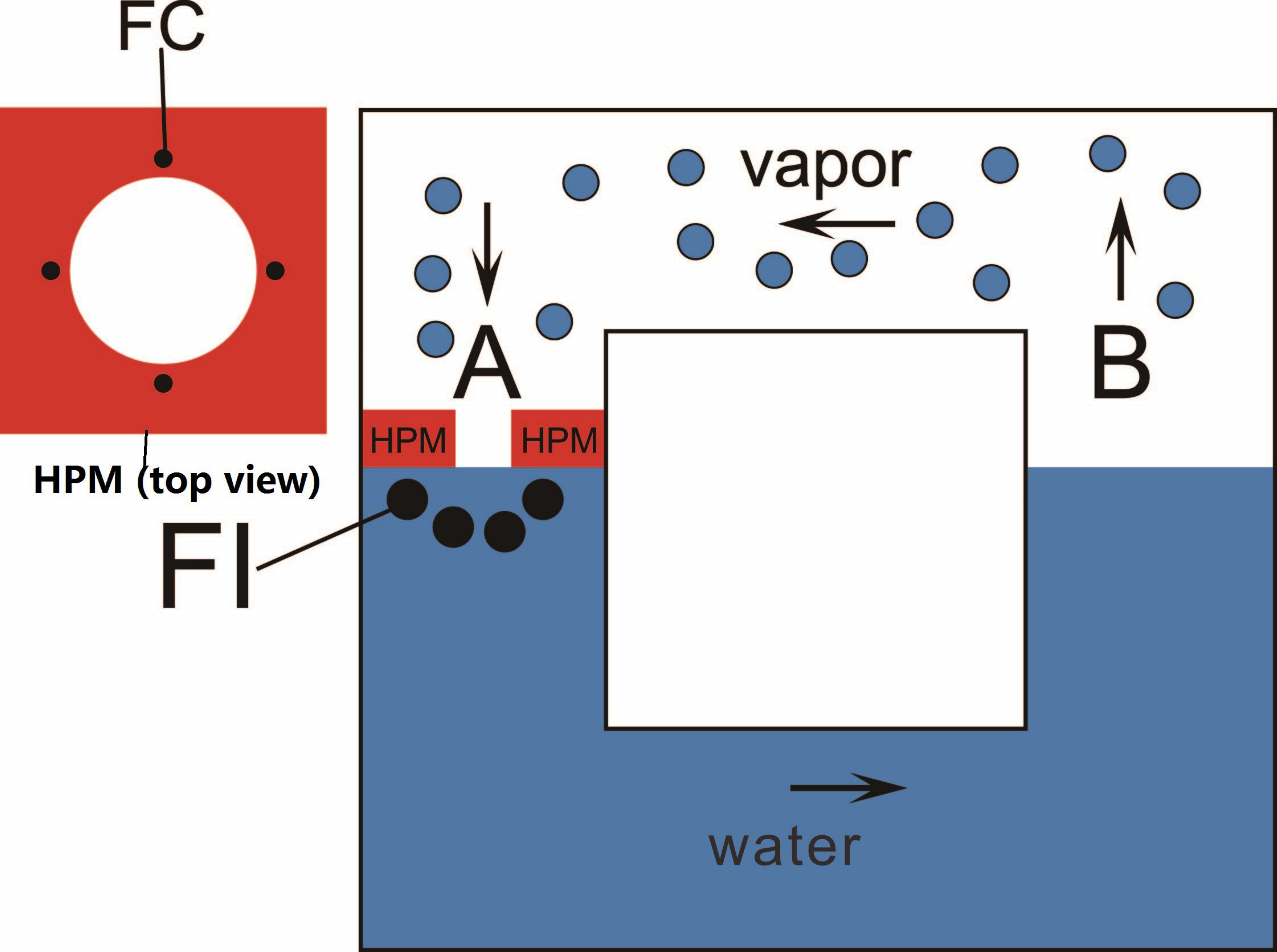}
    \caption{The real physical system underlying the simulation model.
    }
    \label{fig:s1}
\end{figure}

The simulation model is inspired by the real physical system schematically shown in Fig.~\ref{fig:s1}. A certain amount of water is contained in a closed circulation container, forming two liquid–vapor interfaces denoted as surfaces A and B. A hydrophobic porous membrane (HPM) carrying permanent electrostatic charges are placed above surface A. These fixed charges (FC) attract free ions (FI) dissolved in water, such that the ion concentration at liquid surface A is consistently higher than that at surface B.

As demonstrated by our simulations, this ion-concentration asymmetry can continuously induce a directed vapor flux from surface B to surface A. After condensation at surface A, the liquid water flows back toward surface B, forming a closed circulation loop, as indicated by the arrows in Fig.~\ref{fig:s1}.

\section{More details about simulation settings}\label{supp_sec:2}
The hydrophobic porous membrane (HPM) consists of 147 nitrogen molecules, each carrying zero net charge. Water molecules are modeled using the TIP4P model~\cite{S1}, and are therefore polar. The interaction between nitrogen atoms and water molecules is weak, involving only van der Waals forces, so water does not wet the membrane. In contrast, water–water interactions are strong, and ion–water interactions are even stronger due to the presence of both Coulomb and van der Waals interactions.

Energy minimization is performed prior to each simulation. Molecular dynamics simulations are carried out using the leap-frog algorithm (the default integrator in GROMACS) with a time step of 1 fs. Trajectories are recorded every 1 ps. The electrostatic interactions are computed using the particle mesh Ewald method with a real space spherical cutoff of 1.2 nm. Lennard–Jones interactions are also truncated at a distance of 1.2 nm. 

All simulations are performed in the NVT ensemble. Temperature is controlled using a Nosé–Hoover thermostat with a coupling time of 0.2 ps, which effectively connects the system to an infinite heat bath at constant temperature. Because the Nosé–Hoover thermostat does not randomize particle velocities~\cite{S2,S3}, it is well suited for studying transport properties.

Although the model can in principle operate over a wide temperature range, simulations are conducted at 370 K rather than room temperature. This choice is motivated by sampling efficiency: the vapor pressure of water at 370 K is more than 25 times higher than that at 300 K, which significantly reduces the simulation time required to collect sufficient vapor-phase statistics. The simulations are conducted with Gromacs and the force field is Opls\_aa.

\section{More details about simulation settings}\label{supp_sec:3}
In all simulations, the fixed charges on the HPM pore wall always carry the opposite sign to that of the free ions in the water slab, ensuring electrostatic attraction between them. Simulations B1–B6 employ different ionic species as free ions to test the generality of the mechanism.

In every case, free ions accumulate near liquid surface A due to electrostatic attraction, leading to an ion concentration difference between surfaces A and B. The results are summarized in Table~\ref{tab:s1}. In all simulations, the NNWM values are positive and increase with time, demonstrating that the model robustly induces a directed flux in the same direction irrespective of whether the free ions are positively or negatively charged.

\begin{table}[htbp]
	\centering
	\setlength{\tabcolsep}{12pt}   
	\renewcommand{\thetable}{S1}
	\caption{Statistical summary of simulations B1$\sim$B6.}
	\label{tab:s1}
	\begin{tabular}{lcccccc}
		\toprule
		Sim-ID & Type-FC & Type-FI & Temp (K) & Time-sim (ns) & V-NNWM & F-rate (ns$^{-1}$) \\
		\midrule
		B1    & $Cl^-$    & $Na^+$    & 370    & 400    & 399    & 0.998 \\
		B1-1  & $Cl^-$    & $Na^+$    & 370    & {4,000} & 4,352  & 1.088 \\
		B2    & $Cl^-$    & $K^+$     & 370    & 400    & 422    & 1.055 \\
		B3    & $Cl^-$    & $Li^+$    & 370    & 400    & 434    & 1.085 \\
		B4    & $Na^+$    & $Cl^-$    & 370    & 400    & 522    & 1.305 \\
		B5    & $Na^+$    & $F^-$     & 370    & 400    & 360    & 0.9   \\
		B6    & $Na^+$    & $Br^-$    & 370    & 400    & 571    & 1.43  \\
		\bottomrule
	\end{tabular}
\end{table}

Notes: Sim-ID denotes simulation identifier; Type-FC the type of fixed charges; Type-FI the type of free ions; Temp the thermostat temperature; Time-sim the total simulation time; V-NNWM the final NNWM value; F-rate the average flux rate (water molecules/ns).

The overall dynamical behavior is similar across all simulations. Here we focus on simulation B1 as a representative example. Four $Cl^-$ ions are used as fixed charges and four $Na^+$ ions as free ions. The thickness of HPM extends from 8.3 to 9.3 along the $z$ axis. During the initial few picoseconds, $Na^+$ ions migrate toward the HPM pore, dragging dozens of water molecules into the pore region. Statistical analysis shows that the electric field of the fixed charges effectively confines $Na^+$ ions near liquid surface A. The probability of the $Na^+$ ions appearing above $z = 8.3$ is 66.26\%, meaning that the free ions appear in the pore region most of the time.

The system reaches thermal equilibrium within the first few picoseconds. Throughout the simulation, the average number of vapor molecules is approximately 1.58. By the end of the simulation, NNWM reaches 399, confirming the presence of a persistent directed vapor flux.

\begin{figure}[t]
    \centering
    \includegraphics[width = 0.5\linewidth]{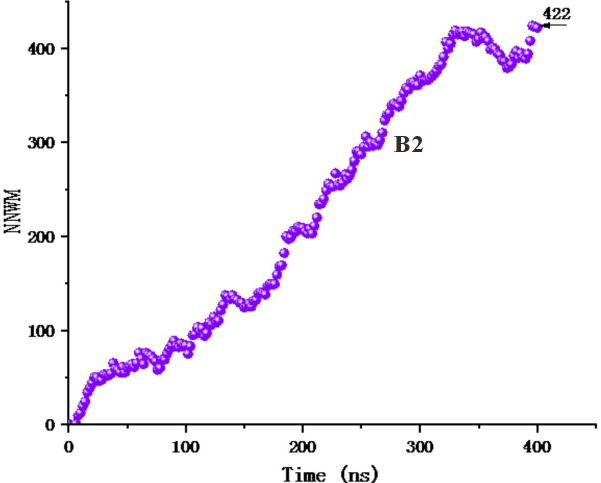}
    \caption{Variation of NNWM value as a function of time in simulation B2.
    }
    \label{fig:s2}
\end{figure}

\begin{figure}[t]
    \centering
    \includegraphics[width = 0.5\linewidth]{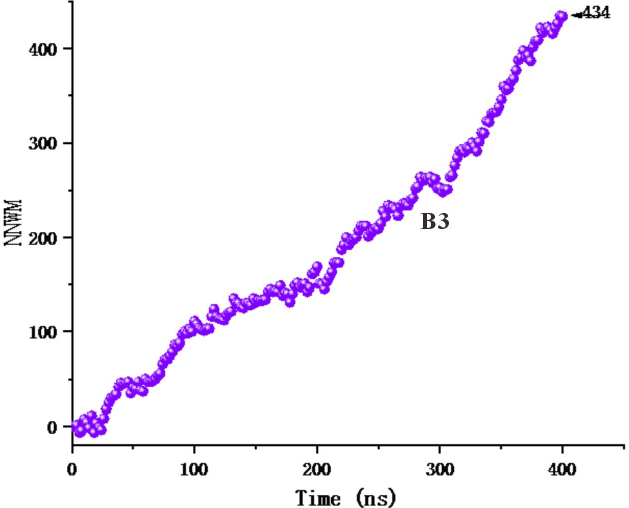}
    \caption{Variation of NNWM value as a function of time in simulation B3.
    }
    \label{fig:s3}
\end{figure}

The line charts of only two additional examples are shown here for discussion. Figs.~\ref{fig:s2} and~\ref{fig:s3} present the NNWM time series for simulations B2 and B3, respectively. The observed fluctuations of NNWM arise from several factors: (i) the intrinsic stochasticity of molecular motion; (ii) occasional excursions of one or more free ions away from the pore region, leading to variations in the influence on surface A; and (iii) transient clustering of vapor molecules, which can cause abrupt changes in NNWM when clusters cross the counting plane.

\section{Saturated vapor pressure difference.}\label{supp_sec:4}
\begin{figure}[t]
    \centering
    \includegraphics[width = 0.4\linewidth]{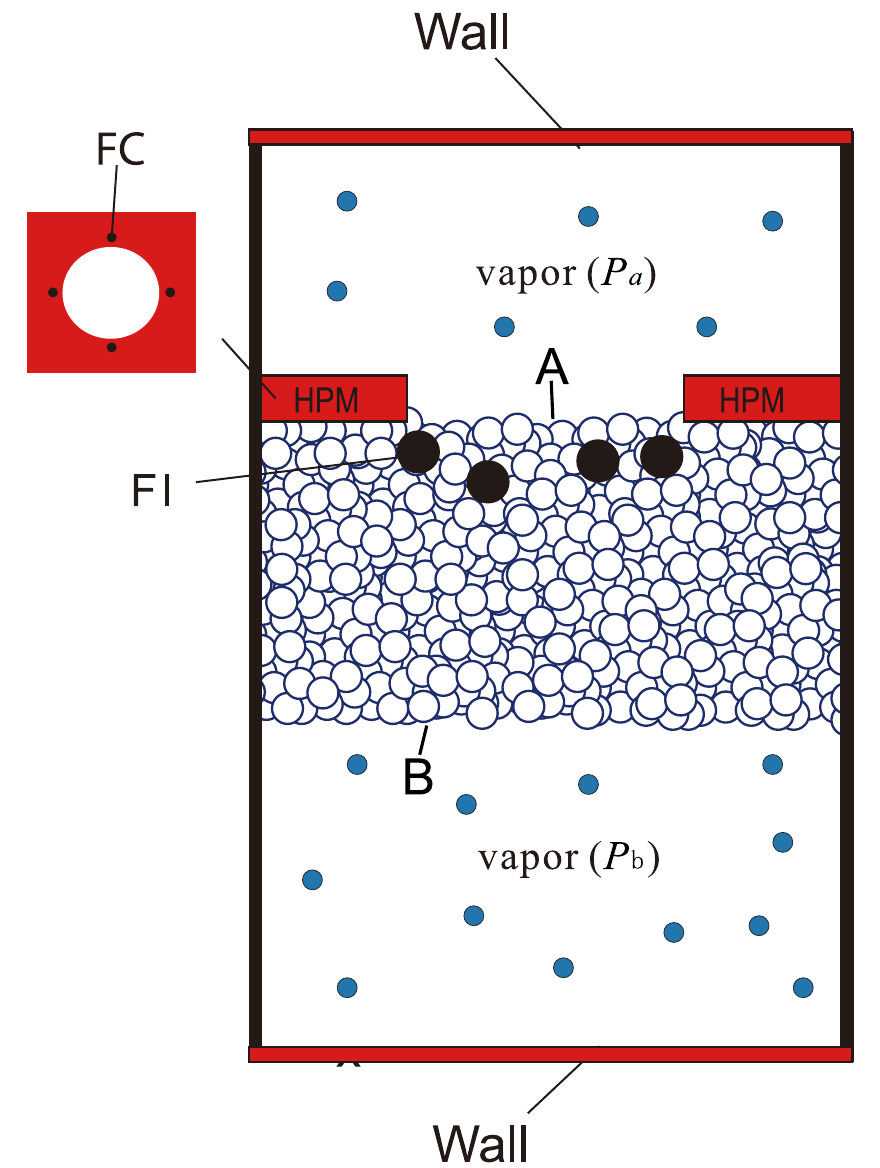}
    \caption{Side view of the model with two separate vapor regions.
    }
    \label{fig:s4}
\end{figure}

Based on the configuration and settings of simulation B1, a virtual wall is introduced at the z-boundary, as shown in Fig.~\ref{fig:s4}, preventing vapor molecules from crossing the periodic boundary and ensuring that the two vapor regions A and B are independent.

The simulation runs for 1,000 ns. Statistical analysis based on one million frames shows that the vapor pressures of two vapor regions A and B are 83.872 kPa and 90.945 kPa, respectively. This corresponds to a pressure ratio (  = 1.084) and a pressure difference of 7.073 kPa (equivalent to a 0.72 m water column). Vapor molecules are more frequently observed on vapor region B, consistent with the lower potential barrier at liquid surface B.

\section{Continuity of the directed flux.}\label{supp_sec:5}
\begin{figure}[t]
    \centering
    \includegraphics[width = 0.5\linewidth]{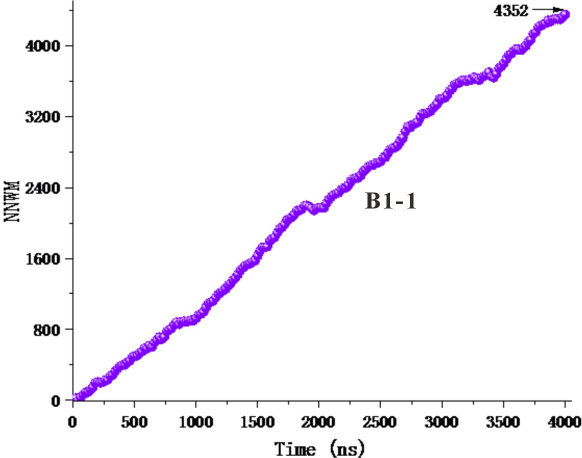}
    \caption{Variation of NNWM value as a function of time in simulation B1-1.
    }
    \label{fig:s5}
\end{figure}

Simulation B1 is extended to 4,000 ns. By the end of the simulation, NNWM reaches 4,352 (B1-1 in Supplementary Table I), which is substantial compared to the total number of water molecules (950). As shown in Fig.~\ref{fig:s5}, the NNWM increases linearly with time and exhibits no sign of decay, indicating a stable steady-state flux.
No free ions escape from surface A, and all condensed water molecules reintegrate into the bulk liquid, preserving the integrity of the water slab. One might suspect that the observed flux originates from residual nonequilibrium effects associated with the initial condition. However, the sufficiently long simulation time rules out this possibility. The persistent trend demonstrates that the directed transport is an intrinsic property of the model rather than a transient relaxation process.

\section{Engineering implementation and optimization.}\label{supp_sec:6}
The polarity and density of surface charges in nanochannels can be tailored by various fabrication techniques~\cite{S4}, such as self-assembled monolayers and surface grafting~\cite{S5}, surface-initiated controlled radical polymerization (SI-CRP)~\cite{S6}, and irradiation-induced functionalization~\cite{S7}. Jiang et al. reported that nanochannels fabricated from mica can adsorb a large number of cations on their surfaces~\cite{S1}, with a packing density close to the 2D dense packing limit. These techniques enable the creation of permanent charged sites on the inner wall of the HPM pore, which is a prerequisite for sustaining directed transport in our model. In addition, electric-field modulation of surface charges~\cite{S9} and optical modulation~\cite{S10} have already been implemented in nanofluidic devices, raising the possibility of developing our model into a new class of externally controllable fluidic devices.

\begin{figure}[t]
    \centering
    \includegraphics[width = 0.5\linewidth]{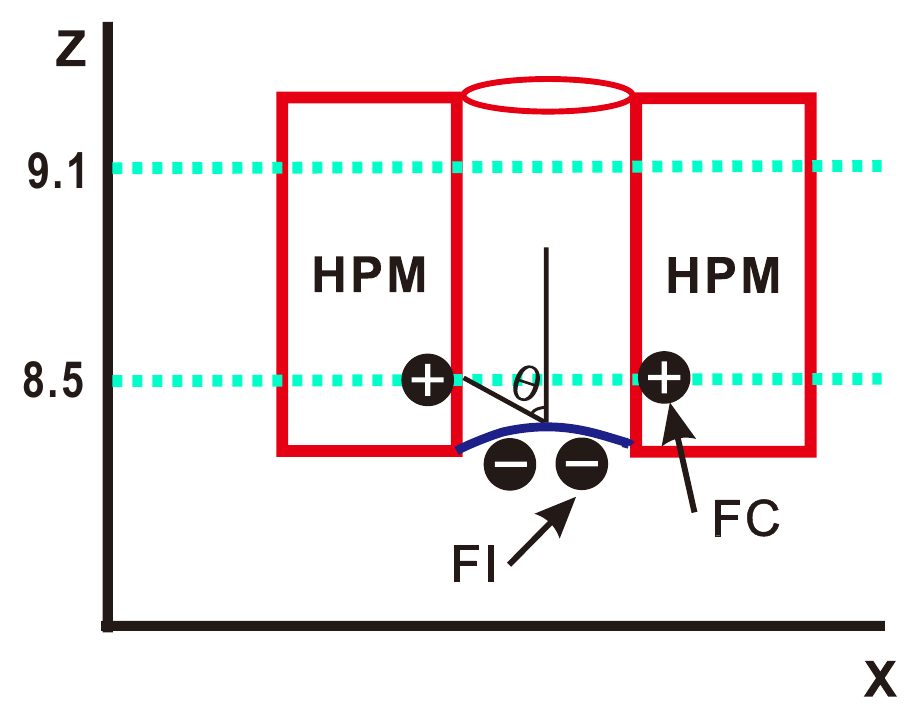}
    \caption{Position variation of fixed charges (sideview). (Note: the numbers of free ions and fixed charges and model size are not drawn to scale.)
    }
    \label{fig:s6}
\end{figure}

Certain control parameters have a pronounced influence on the rectification efficiency. As shown in Fig.~\ref{fig:s6}, the angle   is defined as the angle between the force direction exerted by the fixed charges on free ions and the normal to the liquid surface. Compared with B1, simulations B7–B10 differ solely in the position of the fixed charges along the z axis (see the results summarized in Table~\ref{tab:s2}). The rectification efficiency increases remarkably as the fixed charges are positioned higher along the $z$ axis. As   decreases, the attractive force acting on the free ions acquires a larger component perpendicular to the liquid surface, thereby increasing the probability for free ions to appear at the liquid surface and enhancing their influence on it.
\begin{table}[htbp]
  \centering
  \setlength{\tabcolsep}{12pt}
  \renewcommand{\arraystretch}{1.5}
  \renewcommand{\thetable}{S2}  
  \caption{Results of simulations with fixed charges at different positions.}
  \label{tab:s2}
  \begin{tabular}{lccccc}
    \toprule
    Sim-ID & Plane-FC & Time-sim (ns) & V-NNWM & F-rate (ns$^{-1}$) \\
    \midrule
    B7  & z=8.3 & 400 & 296  & 0.74   \\
    B1  & z=8.5 & 400 & 399  & 0.998  \\
    B8  & z=8.7 & 400 & 548  & 1.37   \\
    B9  & z=8.9 & 400 & 759  & 1.898  \\
    B10 & z=9.1 & 400 & 965  & 2.413  \\
    \bottomrule
  \end{tabular}
\end{table}

Notes: Plane-FC denotes the xy plane at a given z position where the fixed charges are located. Sim-ID, Time-sim, V-NNWM, and F-rate are defined as in Table~\ref{tab:s1}.

However, pore size presents a significant challenge. Radial distribution functions show that the peak probability of free ions around fixed charges occurs at approximately 0.27 nm. Free ions predominantly affect the first hydration shell; although the second hydration shell still experiences the electrostatic field, its influence is substantially attenuated~\cite{S11}. Consequently, the impact of free ions is confined to an extremely narrow region adjacent to the pore wall, implying that the HPM pore diameter must be very small for practical implementations.

Data supporting the findings of this study are available from the corresponding author upon reasonable request.